\documentclass[12pt]{article}
\usepackage{amsmath}
\usepackage{amssymb}
\begin{document}

\begin{center}
{\bf CASIMIR FRICTION  IN TERMS OF MOVING HARMONIC OSCILLATORS: EQUIVALENCE BETWEEN TWO DIFFERENT FORMULATIONS}

\vspace{1cm}
Johan S. H{\o}ye\footnote{johan.hoye@ntnu.no}

\bigskip
Department of Physics, Norwegian University of Science and Technology, N-7491 Trondheim, Norway

\bigskip
Iver Brevik\footnote{iver.h.brevik@ntnu.no}

\bigskip
Department of Energy and Process Engineering, Norwegian University of Science and Technology, N-7491 Trondheim, Norway

\bigskip

\end{center}

\begin{abstract}

The Casimir friction problem can be dealt with in a simplified way
by considering two harmonic oscillators moving with constant
relative velocity. Recently we calculated the energy dissipation
$\Delta E$ for such a case, [EPL {\bf 91}, 60003 (2010); Europ.
Phys. J. D {\bf 61}, 335 (2011)]. A recent study of Barton [New J. Phys.
{\bf 12}, 113044 (2010)] seemingly leads to a
different result for the dissipation. If such a discrepancy really were true, it would imply a delicate difficulty for the basic theory of Casimir friction. In the present note we show that
 the expressions for $\Delta E$ are in fact
physically equivalent, at $T=0$.

 \end{abstract}


\section{Introduction}

Noncontact friction caused by electromagnetic fluctuations - also called van der Waals or Casimir friction - has recently become a topic of considerable interest in the Casimir community. Early studies in this direction were made by Teodorovich \cite{teodorovich78},   Levitov \cite{levitov89}, and by the present authors \cite{hoye92,hoye93}. More recent papers on Casimir friction, viewing the effect from various perspectives, can be found in Refs.~\cite{pendry97,pendry98,volokitin07,volokitin08,philbin09,philbin09A,dedkov08,dedkov08A,dedkov10,pendry10,hoye10,hoye10A,barton10,barton10A}. Even the somewhat exotic topic of noncontact friction between a graphene sheet and a SiO$_2$ substrate has been studied most recently \cite{volokitin11}.

However, in spite of these developments the theory of Casimir friction has left some subtle issues, even on the fundamental level.  In the present note we shall focus attention on one specific issue of this sort, namely the noncontact friction force and the associated energy dissipation when, instead of two slabs, {\it two harmonic oscillators} are assumed to move relative to each other with a constant nonrelativistic velocity. Perhaps counter-intuitively, such a  seemingly very simple model possesses quite  nontrivial properties. We actually considered this kind of Casimir system in two recent papers \cite{hoye10,hoye10A}, whose main results were that at finite temperature $T$ there exists a finite friction force between the oscillators. At $T=0$ we found the friction force to be zero; this result, however, being due to the assumption about a very slowly varying coupling. We will reconsider this briefly at the end of the next section. For more realistic cases involving a rapidly varying coupling, a finite friction force would be the case also when $T=0$.

Here it can be remarked that the static van der Waals force between the oscillators may affect the trajectory of the particles when they are close. However, to avoid the additional complications of a realistic two-body scattering process the relative motion is assumed to be rigidly guided in a prescribed way. Equivalently, the masses of the oscillators can be regarded as infinite.

In the present paper we shall assume $T=0$. The investigation below was inspired by two recent papers of Barton \cite{barton10,barton10A}, the first of which was dealing with a two-oscillator model, thus in essence the same microscopic model as ours. Barton analyzed the system using quantum mechanical perturbation theory. The striking point is that the expressions he obtained are seemingly quite different from those we obtained in Ref.~\cite{hoye10}. In Barton's own words (\cite{barton10}, Sect. 3) "..in view of the manifold current controversies about quantum-governed frictional force generally, it seems well worth exploring whether such differences reflect substantiate disagreement or only a confusion of terms".

Our present investigation is  a follow-up of Barton's suggestion. We intend to demonstrate explicitly that the obtained expressions for the dissipated Casimir energy are in fact  in agreement with each other, thus a reassuring result.
To facilitate and shorten this comparison our recent results obtained in Ref.~\cite{hoye10A} are utilized. In Ref.~\cite{hoye10} the energy dissipated was obtained via a Kubo formalism where the perturbing interaction was integrated together with a response function. In Ref.~\cite{hoye10A}, however, general expressions for time-dependent perturbation theory were obtained, and we were able to show the equivalence to those of Ref.~\cite{hoye10} for arbitrary $T$. Barton considered a specific model of 2 particles art $T=0$, and below we will show that his expression for the dissipation agrees with the one of Ref.~\cite{hoye10A} and thus with the one of Ref.~\cite{hoye10}.

\section{Comparison at $T=0$}

Let  the dissipated Casimir energy for the two-oscillator system be called   $\Delta E$. The oscillators are assumed to have the same eigenfrequency $\omega$ and the same mass $m$. They interact via a time-dependent Hamiltonian
\begin{equation}
H_{\rm int}=\frac{e^2}{s^3}\,y_1y_2, \label{1}
\end{equation}
(Gaussian units assumed).  Here $e$ is the elementary charge, $y_1$ and $y_2$ the oscillator coordinates, and ${\bf s =s}(t)$ is the  vectorial distance between the mass centers, varying with time because of the relative motion of the oscillators.  Introducing new coordinates
\begin{equation}
y_\pm =\frac{y_1 \pm y_2}{\sqrt 2}, \label{2}
\end{equation}
one can write the interaction Hamiltonian as
\[ H_{\rm int\pm}=H_{\rm int+}+H_{\rm int-}, \]
\begin{equation}
H_{\rm int\pm}=\pm \frac{1}{2}q\,y_\pm^2, \quad q=\frac{e^2}{s^3}. \label{3}
\end{equation}
By use of time-dependent perturbation theory the total energy dissipated is then found to be
\[ \Delta E=2\times 2\hbar \omega |c(\infty)|^2,  \]
\begin{equation}
c(t)=-\frac{i}{2\hbar}\int_{-\infty}^t dt' q \,\langle 2_+ |y_+^2|0_+\rangle \, e^{2i\omega t'}, \label{4}
\end{equation}
as given by Eqs.~(3.3) and (2.4) respectively, in Ref.~\cite{barton10}. The result (\ref{4}) is valid for temperature $T=0$ where only excitations from the ground state are possible.

To compare Barton's result (\ref{4}) with ours, we first have to evaluate the matrix elements in (\ref{4}). In terms of the creation and annihilation operators we have ($\omega_\pm \rightarrow \omega$ for small perturbations)
\begin{equation}
y_\pm= \sqrt{b}\,(a_\pm+a_\pm^\dagger),\quad b=\frac{\hbar}{2m\omega}. \label{5}
\end{equation}
Then,
\begin{equation}
\langle 2_+ |y_+^2 |0_+ \rangle=b\,\langle 2_+ |{a_+^\dagger}^2 |0_+ \rangle =\sqrt{2} \,b. \label{6}
\end{equation}
Together with Eq.~(\ref{4}) this gives
\[ \Delta E=8\hbar \omega b^2 |I(\infty)|^2, \]
\begin{equation}
I(t)=-\frac{i}{2\hbar}\int_{-\infty}^t dt' q\, e^{2i\omega t'}. \label{7}
\end{equation}
Proceed now to compare this result with those that we derived in Refs.~\cite{hoye10} and \cite{hoye10A}, restricting us here to zero temperature. The results obtained in Refs.~\cite{hoye10} and \cite{hoye10A} had quite different forms, as different methods were used in their derivations. In Ref.~\cite{hoye10A}, however, they were shown to be equivalent. The interaction Hamiltonian was written as $H_{\rm int}=-Aq(t)$, with
\begin{equation}
A=-y_1y_2, \quad  {\rm and} \quad q(t)=q=\frac{e^2}{s^3}. \label{8}
\end{equation}
Thus $A$ is a time-independent operator accounting for the quantum mechanical properties of the system, while $q(t)$ is a classical function of time. At $T=0$ the system is in its ground state with probability equal to one. The perturbation $H_{\rm int}$ can excite each oscillator only to the level lying above the ground state.

From Eq.~(15) in Ref.~\cite{hoye10A} we have for the change in energy
\begin{equation}
\Delta E=\sum_{nm} (E_n-E_m)P_m B_{nm}, \label{9}
\end{equation}
where $E_n$ is the energy in the (unperturbed) eigenstate $n$, and $P_n=|a_n|^2$ is the probability of the system to be in this state. As mentioned, we start from the ground state so that $P_m \rightarrow P_{00}=1$. Further, $B_{nm}=|b_{nm}|^2$ with $b_{nm}$ the transition coefficient between states $m$ and $n$. Then, Eq.~(\ref{9}) reduces to
\begin{equation}
\Delta E=(E_{11}-E_{00})B_{1100}, \label{10}
\end{equation}
where $E_{11}-E_{00}=2\hbar \omega$ is the energy difference between the state $|11\rangle$ where both oscillators are excited to the first level, and the ground state $|00 \rangle$. The coefficient $B_{1100}$ is the transition probability between these two states.

What remains is to calculate $B_{1100}$. From Eq.~(10) in Ref.~\cite{hoye10A} we have
\[ b_{nm}=-\frac{1}{i\hbar}A_{nm}\hat{q}(-\omega_{nm}), \]
\begin{equation}
\hat{q}(\omega)=\int_{-\infty}^\infty q(t)e^{-i\omega t} dt, \label{11}
\end{equation}
the hat denoting Fourier transform. Then
\begin{equation}
\hat{q}(-\omega_{nm})\rightarrow \hat{q}(-\omega_{1100})=\hat{q}(-2\omega)=2i\hbar I(\infty), \label{12}
\end{equation}
with $I(\infty)$ given by Eq.~(\ref{7}). Further,
\[ A_{nm} \rightarrow A_{1100} =\langle 11|-y_1y_2|00\rangle \]
\begin{equation}
=-\langle 1|y_1|0\rangle \langle 1|y_2|0\rangle=-\langle 1|y_1|0\rangle^2=-b\langle 1|a_1^\dagger |0\rangle^2 =-b. \label{13}
\end{equation}
Altogether, when inserted into Eq.~(\ref{10}) we obtain
\[ B_{1100}=\frac{1}{\hbar^2}|A_{1100}|^2\hat{q}(-2\omega)\hat{q}(2\omega)=4b^2|I(\infty)|^2, \]
\begin{equation}
\Delta E=8\hbar \omega b^2|I(\infty)|^2, \label{14}
\end{equation}
which coincides with the result obtained by Barton, Eq.~(\ref{7}) above.

As an additional remark we note that the situation with zero friction at $T=0$ for slowly varying forces can be analyzed in a straightforward way from the equations above. With time-dependent part ($q=0$, $t<0$),
\begin{equation}
q(t)=\gamma te^{-\eta t}, \quad t>0,
\label{15}
\end{equation}
with $\gamma$ a constant, we namely find from Eq.~(\ref{11}) when $\eta \rightarrow 0$
\begin{eqnarray}
\hat q(\omega)&=&\frac{\gamma}{(\eta+i\omega)^2},\\
\hat q(\omega)\hat q(-\omega)&=&\frac{\gamma^2}{\eta^2+\omega^2}\rightarrow \frac{\pi \gamma^2}{2\eta\omega^2} \delta(\omega).
\label{16}
\end{eqnarray}
The explicit definition of the quantity $\gamma$  is given in Eq.~(35) in \cite{hoye10A}; it will not be further needed here. Thus with $\omega\neq 0$, Eq.~(\ref{16})  will give zero for the dissipated energy $\Delta E$.

 [By looking at this in some more detail it would seem that the middle term in Eq.~(\ref{16}), before taking the limit
  $\eta\rightarrow 0$, implies there to be some  dissipation. This is physically an artefact, due to our assumption of an abrupt change of  $q(t)$ at $t=0$. Upon division with  the decay time $1/\eta$ this part will vanish too. Alternatively, one can get rid of  this latter dissipation by choosing $q(t)=t e^{-q|t|}$ for all $t$ by which $\hat q(\omega)=-4i\eta\omega \gamma/(\eta^2+\omega)^2$.]

The equivalence between Barton's results and ours, as anticipated in the Introduction, is therewith shown. We thus hope to have shed light on one of the subtle issues in the  Casimir friction world demonstrating that quite different approaches can lead to the same result.

Finally, the reader is referred to a most recent paper of Barton \cite{barton11}, dealing with the van der Waals friction between two atoms at nonzero temperature.



\bigskip
{\bf Acknowledgment}

\bigskip
I. B. thanks Gabriel Barton for valuable correspondence.

\end{document}